\documentclass[12pt,preprint]{aastex}

\newcommand{\HII}{\hbox{{\rm H}\kern 0.1em{\sc ii}}}
\newcommand{\HI}{\hbox{{\rm H}\kern 0.1em{\sc i}}}

\begin{document}

\title{\large An HST Archival Survey of Feathers in Spiral Galaxies}
\author{\normalsize Misty A. La Vigne, Stuart N. Vogel, and Eve C. Ostriker}
\affil{Department of Astronomy, University of Maryland, College Park, MD 20742}


\begin{abstract}
  
  We present a survey of spiral arm extinction substructure referred
  to as feathers in 223 spiral galaxies using HST WFPC2 images.  The
  sample includes all galaxies in the RC3 catalog
  with $cz$ \(<\) 5000 km s\(^{-1}\), B\(_{T}\) \(<\)~15, {\it i}
  \(<\)~60\degr, and types Sa--Sd with well-exposed broadband WFPC2
  images.  The detection frequency of delineated, periodic feathers in
  this sample is 20\%\ (45 of 223).  This work is consistent with
  Lynds (1970), who concluded that feathers are common in prototypical
  Sc galaxies; we find that feathers are equally common in Sb
  galaxies.  Sb--Sc galaxies without clear evidence for feathers
  either had poorer quality images, or flocculent or complex
  structure.  We did not find clearly defined feathers in any
  Scd--Sd galaxy.  The probability of detecting feathers was
  highest (83\%) for spirals with well-defined primary dust lanes (the
  lanes which line the inner edge of an arm); well-defined primary dust lanes were
  only noted in Sab--Sc galaxies.  The detection frequency of feathers was similar
  in barred and unbarred spirals.  Consistent with earlier work, we
  find that neighboring feathers tend to have similar shapes and pitch
  angles.  Well-defined feathers often emerge from the primary dust
  lane as leading features before they curve to trailing; some are
  quite elongated, extending into the interarm and merging with
  other feathers.  OB associations are often found lining feathers,
  and many feathers transition to the stellar substructures
  known as spurs (Elmegreen 1980).  We find that feathers are
  coincident with interarm filaments strikingly revealed in
  Spitzer 8$\mu$m images.  Comparison with CO 1-0 maps of
  NGC 0628 and NGC 5194 from BIMA SONG shows that feathers
  originate at the primary dust lane coincident with gas surface
  density peaks.  Contrary to the appearance at 8$\mu$m, the
  CO maps show that gas surface density in feathers decreases rapidly
  with distance from the primary dust lane.  Also, we find that the
  spacing between feathers decreases with increasing gas surface
  density; consistent with formation via a gravitational instability.


\end{abstract}


\maketitle

\section{Introduction}
Spiral arms are not smooth, continuous features.  Typically an arm is
composed of many substructures commonly referred to as feathers,
spurs, and branches, which give the arm its patchy, divaricate
appearance.  This substructure appears associated with much of the
star formation in the arm.

\subsection{Previous Observations of Spiral Arm Substructure}\label{previous_obs}

Early reports of spiral arm substructure were based on examination of
photographic plates such as those used for prints in \textit{The
  Hubble Atlas of Galaxies} (Sandage 1961).  Weaver (1970) noted that
the spiral arms of nearby galaxies appear clumpy, irregular, and
mottled on small scales.  
In particular, he noted the presence of ``spurs" (also referred to
as branches or twigs), which appear to originate on the {\it outside}
of the arm, with larger pitch angles than the arm itself.  Spurs extend
from the outside of the arm into the interarm, and are seen in
photographic plates as {\it stellar} features.  Weaver also noted the
presence of {\it dark} material concentrated along the {\it inner} edges
of spiral arms.  He remarked that the outside regions of arms appear to be
made of material drawn out or ``brushed" from the inner edges, and
that this brushed-out structure has a pitch angle typically a factor of
two larger than the arm itself.


At the same time, Lynds (1970) reported a detailed study of dark
nebulae in 17 late-type spirals using photographic plates from Mt.
Wilson and Palomar Observatories.  She commented on the well-known
strong dust lanes along the inner edge of the arms, which she termed
``primary dust lanes" (hereafter abbreviated PDLs), and noted the
presence of thin dust lanes with large pitch angles cutting {\it
  across} the luminous arms.  She called these features ``feathers",
and pointed out that these extinction features become mostly
undetectable outside the luminous arm presumably owing to the absence
of a sufficiently bright background.  She also emphasized that bright
{\HII} regions are typically near or embedded in dust lanes.  Later,
Piddington (1973) noted that interarm stellar features called
spurs by Weaver are often found associated with the extinction
features cutting across luminous arms called feathers by Lynds,
and suggested that the two are related.


Subsequently, Elmegreen (1980) studied seven spiral galaxies to
investigate the properties of spurs, which as noted above are
observable as stellar features.  Elmegreen was able to identify two to
six well-delineated spurs in each of her galaxies, with lengths
ranging from one to five kpc; her spurs are generally located in the
outer parts of the luminous disk.  She observed that spurs are always
located on the outside of spiral arms, have pitch angles equal to or
greater than that of the originating arm, and commonly occur in
pairs or groups with the spurs oriented roughly parallel to one another.
Elmegreen also determined that on average spurs have pitch angles of
roughly 50\degr\ with respect to spiral arms, which is comparable to
the average pitch angles of the feathers measured by Lynds (1970).  Noting the
similarity between the pitch angles of feathers and spurs, Elmegreen
added further observational support to Piddington's (1973) suggestion
that spurs and feathers have a common origin.

\subsection{Theoretical Studies of Substructure}\label{theories}

Motivated to explain the formation of observed spurs,
branches, and feathers in spiral galaxies, Balbus (1988, hereafter
B88) conducted a local gas dynamical stability analysis of a
single-fluid polytropic flow through spiral arm potentials, following
the linear evolution of self-gravitating perturbations.  In
his analysis, the background gaseous surface density profile
representing the arm has an arbitrary spatial form, and the
differentially rotating, expanding background flow is consistent with
this profile.  Balbus investigated all wavenumber directions in the
plane of the disk and modeled the spiral arms as tightly wound, with
no magnetic field.  B88 found that there are two preferred directions
of growth in spiral arm flow: initial wavefronts roughly along the
spiral arm, or perpendicular to it.  The rate of growth in both
directions depends on the properties of the underlying flow.  In
certain regimes, growth of instabilities leads to fragmentation
parallel to the arms, observed as a thickening of the arms.  In other
situations, growth of initially leading wavenumbers results in
branches, feathers, or spurs.  For wavefronts initially perpendicular
to the arm, as the gas moves into the interarm region, the flow expansion
and shear (which increases downstream from the arm) shapes and stretches 
small-scale structure into the familiar large scale trailing `spur'
shapes that are observed. 


B88 also suggested that a two-dimensional lattice of small-scale structure
can develop when the two dominant modes of growth intersect.    
In \textit{The Hubble Atlas of Galaxies}
(Sandage 1961), B88 found that some barred spiral galaxies, in
particular the western arm of NGC 1300, seemed to exhibit a lattice
structure of {\HII} regions.  However, he found no examples of such
structure in unbarred galaxies.

Kim and Ostriker (2002, hereafter KO) extended the local models of B88
by including the effects of non-linearity and magnetic fields.  KO
conducted local, two-dimensional, time-dependent, magneto-hydrodynamic (MHD)
simulations of self-gravitating, differentially rotating, razor thin
disks of gas.  Their models followed the formation and fragmentation
of ``gaseous spurs'' as the flow passes through spiral arm potentials.
They found that local substructures are created via the magneto-Jeans
instability (MJI); as the background flow passes through the spiral
pattern it is shocked and compressed until it becomes Jeans unstable,
at which point gravity, aided by magnetic forces, begins to create
alternating compressed and rarefied regions along an arm.  The
magnetic effects aid the formation of compressed, self-gravitating
complexes because magnetic tension forces oppose the Coriolis forces
that would otherwise stabilize the flow, helping to transfer angular
momentum out of growing condensations.  KO found that
the gaseous spurs fragment into clumps within which star formation
could commence.  They suggest that these clumps could be the precursors
of bright {\HII} regions that jut from the outside of spiral arms
inside corotation.  

From their simulations, KO put together observable
statistics of their gaseous spurs.  Most
potentially comparable to observations is the spacing of these
features along a spiral arm, which they find ranges from 2--5 times the
local Jeans wavelength (at the spiral arm density peak) and
corresponds to a spacing of approximately 750 pc on average.  KO also
proposed that the shape and location of gaseous spurs within a
spiral arm may potentially be used observationally to determine the
spiral pattern speed of the arm.  Very recently, Kim \& Ostriker (2006)
have extended their thin disk simulations to three dimensions, also making
comparison to two-dimensional ``thick disk'' models.  The results they find are
overall consistent with the conclusions of KO, with the difference
that spur spacings increase by a factor \(\sim2\) due to the dilution of self gravity when
the disk thickness increases.

Chakrabarti, Laughlin, and Shu (2003, hereafter CLS) studied the
response of a thin, self-gravitating, singular isothermal gaseous disk
to rigidly rotating spiral potentials, specifically focusing on the
effects of ultraharmonic resonances by choosing parameters that minimize swing 
amplification.
In simulations with
low \(Q_{g}\), CLS found growth of spiral arm ``branches'' (which they define
as trailing bifurcations of the main spiral arms).  
Long-term simulations with high \(Q_{g}\) exhibited
the growth of stubby leading structures (referred to as ``spurs'' by CLS; this is however inconsistent
with terminology of other authors).  


Wada and Koda (2004, hereafter WK) performed two-dimensional, time dependent,
global hydrodynamical simulations of a thin, isothermal,
non-self-gravitating disk of gas in tightly and loosely wound, rigidly
rotating spiral potentials.  WK's model rotation profiles included
both differentially rotating and rigidly rotating cases. They found
that the spiral shock front is stable when the gas is modeled with a
flat rotation curve and unstable when modeled with a rising rotation
curve.  In their models, they found the stability of the shock front
is also dependent upon the pitch angle of the spiral arms: stable if
\(i\leq10\degr\), unstable if \(i\geq10\degr\).  
In the unstable models of WK, strong shocks with arm-interarm density 
ratio \(\sim100\) become unstable by rippling.
WK attribute this ``wiggle'' instability to Kelvin-Helmholtz (K-H)
modes involving the strong velocity shear behind the shock.
Over time, the instabilities become non-linear forming what WK refer
to as ``spurs'' in the interarm regions.  The spurs are
quasi-regularly spaced, approximately 100-200 pc apart; the authors do not state, however, 
how spacing depends on model parameters.   Due to the
shape of the rotation curve, the spurs formed in the WK simulations
are curved near the arm in the opposite sense (i.e. trailing then leading)
to the gaseous spurs produced in
the KO simulations and the small-scale structure predicted in B88's
analysis.  

Dobbs and Bonnell (2006, hereafter DB) used three-dimensional SPH
simulations to study the response of isothermal, non-self-gravitating
gaseous disk models with varying temperatures
 to a four armed rigidly-rotating spiral potential.  DB find that the
temperature of the disk has a crucial effect on the growth of spiral arm
substructure.  In the lower temperature (T \(<10^{3}\) K) models of DB,
the initially smooth arms become clumpy, and then the clumps are sheared into 
trailing features as they return to interarm regions.  The shapes of the interarm
features found by DB are similar to those of KO, reflecting the flat rotation curve 
they adopt.  DB use somewhat nonstandard terminology in describing their results; they 
refer to the portions of interarm extinction features adjoining arms as ``spurs'', and the portions further downstream as ``feathers''.  In their T = 50 K model, which has arm-interarm contrast \(\sim50\), the spacing of DB's spurs are \(\approx 700\) pc;  DB also do not state, however, how spacing depends on model parameters.


\subsection{Discussion of Substructure Nomenclature}

As is evident from the above summaries, the terms ``spurs'', ``feathers'',
and ``branches'' have been used in many ways.  In this paper, we adopt
the definitions from the initial, observational papers:

\textit{feathers} -- thin dust lanes or extinction features that
extend outward at a large angle from the \textit{primary dust lane}
(PDL) which lines the inner side of the arm, cutting across the outer
bright part of the spiral arm (Lynds 1970).

\textit{spurs} -- bright chains of OB associations and {\HII} regions that
jut at a large angle from the spiral arm into the interarm (Weaver 1970, Elmegreen 1980).

\textit{branches} -- divarications of a spiral arm that lend to the
overall spiral structure (Elmegreen 1980).

 \subsection{A New Survey of Spiral Arm Feathers}
 Feathers are of considerable interest because there are both
 observational (Lynds 1970, Piddington 1973, Elmegreen 1980, Scoville
 et al. 2001) and
 theoretical (B88, KO) reasons to associate them with a significant
 portion of star formation in spiral galaxies.  Also, they may provide
 information on basic physical conditions in spiral arms, such as the
 mean gaseous surface density and magnetic field strength (KO).  At
 least in principle, they may also be used to deduce the spiral
 pattern speed and details of the gas flow through spiral arms (KO,
 CLS).  Lastly, they are a striking characteristic of prototypical
 grand design spiral galaxies (e.g. M51, Beckwith 2005) and may
 therefore tell us something about the evolutionary and environmental
 aspects of spiral structure.  A complete description of spiral arms
 should include a characterization of the frequency and properties of
 such feathers, and a complete model of spiral structure should
 explain their dynamical origins.
 
 The goals of this survey include: confirming that feathers are a common
 feature of spiral galaxies, evaluating their frequency and
 characteristics, and determining the types of spiral galaxies in which
 feathers occur (barred or unbarred, early or late-type, grand design
 or flocculent).  We aim to identify where feathers are located: in the inner or
 outer disk, on the inside of spiral arms or the outside?  In addition,
 we investigate the relation of the spacing of feathers to the gas surface
 density of a galaxy.

 \section{Sample Selection}

 Our study was motivated by the WFPC2 Hubble Heritage image of M51
 (Scoville et al. 2001), which revealed remarkable feather structure
 particularly in the inner galaxy.  Since similar structure is
 difficult to discern in a casual inspection of M51 and other galaxies
 in \textit{The Hubble Atlas of Galaxies} (Sandage 1961), we concluded
 that high angular resolution is important and sought to explore the
 Hubble Space Telescope (HST) archives for additional examples of
 galaxies with clearly delineated feathers.

 The HST archive contains thousands of galaxy images.
  To narrow the list, we initially used the
 Elmegreen \& Elmegreen (1987; hereafter EE) sample of 762 spirals, each
 classified as one of nine arm types ranging from ``chaotic,
 fragmented, unsymmetric'' to ``two long symmetric arms dominating the
 optical disk''.  The EE sample might appear ideal, since we expected
 galaxies with well-defined spiral arms to be the best targets.
 However, many galaxies classified as grand design by EE have poorly-defined 
 arms in HST images.  Conversely, many galaxies classified by
 EE as having poorly-defined arms have well-defined arms in HST images.
 The EE classifications appear to be
 determined primarily by the arm structure outside the inner 1\arcmin\ 
 of galaxies, beyond the region typically imaged by HST.  This may be because
 in the Palomar Sky Survey images used by EE the inner arms are
 often saturated and the outer 
 arms dominate the spiral structure.
 We conclude that EE arm classes do not provide
 reliable descriptions of arm type in the inner 1\arcmin\ of many nearby
 galaxies.

 We therefore expanded our search using the \emph{Third Reference
   Catalogue of Bright Galaxies} (RC3).  This increased the sample size by
 including galaxies at more southern declinations not available in the
 EE sample, and allowed us to specify criteria optimized for
 identification of substructures.  Based on the absence of clear
 examples of feathers in the photographic reproductions of ground-based
 images in the Hubble
 Atlas, we decided that distance to a galaxy should be the primary
 criterion for inclusion in the sample.

 \subsection{A Distance-Limited Sample}

 Using Vizier, we selected all galaxies in the RC3 catalog with de
 Vaucouleurs type Sa--Sd, \(cz<\) 5000 km
 s\(^{-1}\), {\it i} \(<\) 60\degr, and B\(_{T}\)~\(<\)~15.  Our sample
 excludes galaxies with type Sdm or later because these typically
 lack well-delineated spiral arms.  

 The
 recessional velocity was limited to \(<\) 5000 km s\(^{-1}\) to ensure
adequate linear resolution to resolve typical dust lanes and any substructure that 
may be present.  Assuming H\(_o\) = 71 \(\rm km~ s^{-1}~ Mpc^{-1}\), this redshift 
corresponds to
a linear resolution of 34 pc for the 0.1\arcsec\ resolution of the three
Wide Field Cameras.
The limit for
 inclination was set to 60\degr\ because at higher inclinations a
 galaxy may be too inclined to clearly discern structure.  The
 B\(_{T}\) magnitude was limited to galaxies brighter than 15th
 magnitude to eliminate small galaxies or low surface brightness
 galaxies, neither of which would be well suited for examination of
 extinction substructure.  (In fact, few galaxies
 in the RC3 catalog at \(cz<\) 5000 km s\(^{-1}\) have
 B\(_{T}\) \(>\) 15, so this criterion did not significantly affect the
 sample size.) The total number of galaxies that meet these criteria is
 630.

 Ideally, for detection of
 feathers (extinction substructure), we need images taken with high
 resolution, a large field of view (FOV), and good sensitivity.  These
 criteria are best satisfied by the HST ACS (Advanced Camera for Surveys)
 instrument.  However, many of
 the ACS images were still proprietary at the time of this study.  We therefore used
 images taken with WFPC2 (Wide Field/Planetary Camera 2),
 which also satisfies the above criteria.

 Since good signal-to-noise ratios are essential, 
 we restricted the sample to images with
 exposure time \(>\) 60 seconds obtained with
 wideband (W) filters.  All images taken with filters
 shorter than 4311 \AA\ and longer than 8012 \AA\ were
 excluded due to the poor sensitivity of WFPC2 at shorter wavelengths,
 and because extinction features are less prominent at longer optical
 wavelengths.  Lastly,
 observations of galaxies where the FOV did not cover a spiral arm were
 excluded from the sample.  We found that 223 of the 630 RC3 galaxies
 had WFPC2 images meeting our criteria.


 \section{Data Reduction}
 
 We obtained from the HST archive calibrated, full
 resolution FITS files for images of all 
 223 galaxies which met the selection criteria described above.  
 We then mosaiked the images and removed cosmic
 rays using standard procedures outlined in the HST handbook.
 
 We selected images of 21 of the galaxies with well-defined substructure to
present as prototypes (see Figures 1--14).  Typically the original images
 vary significantly in brightness between the nucleus and outer
 regions, making it difficult to see substructure in different parts
 of the galaxy in a single image stretch.  Therefore, images shown in
 the figures were further processed by subtracting a radial profile
 derived from the median value in annuli deprojected assuming the RC3
 inclination and using nuclear positions obtained from the NASA/IPAC
 Extragalactic Database (NED).

 \section{Results}

 \subsection{General Characteristics of Feathers}\label{char}

 We illustrate many of the general characteristics of classic feathers
 using HST observations of NGC 5194 and NGC 0628. Images are shown in
 Figures~\ref{m51.overlay} and~\ref{n0628.overlay} with thin lines to
 mark identified feathers, and in Figures~\ref{m51.no.overlay}
 and~\ref{n0628.no.overlay} without the lines.  Each arm exhibits dark,
 delineated extinction features, which emerge from the primary dust
 lane (PDL) that runs along the inside part of the arm, traverse across the luminous arm, 
 and often extend into the interarm.


 Typically, close inspection shows that well-delineated feathers
 emerge from the PDL as leading features and then gradually curve to
 become trailing features further downstream, on the outer edges of
 the spiral arms.  Feathers that are near each other generally have
 similar pitch angles and curvature, as noted by Lynds (1970).  Many
 are dotted with bright knots of star formation or {\HII} regions,
 especially near the PDLs and the beginning of the interarm regions.

 Many of the feathers extend well into the interarm regions.  Several
 coalesce and can be traced through more than 180\degr.  The
 farthest-reaching feathers appear to merge with the PDL
 of the next arm, as is seen in both arms of NGC 0628 and NGC 5194.

 \subsection{Frequency of Feather Detections}\label{frequency}

 Motivated by the observational and theoretical studies reviewed in
 \S\,\ref{previous_obs} and \S\,\ref{theories}, we considered a galaxy to
 have well-defined, classic feathers if it exhibited extinction
 substructure meeting the following criteria: multiple dust lanes with
 approximately regular spacing emerging from a PDL at roughly similar
 large angles, frequently associated with star formation regions.
 
 Feathers were clearly detected in 45 of the 223 (20\%) galaxies
 examined.  To investigate which galaxies have feathers, 
 we classified all 223 galaxies according to how well the spiral arm PDLs
 are delineated, using the HST images.  PDL delineation was categorized from ``poor'' (see NGC
 5055, Figure~\ref{n5055.n5985}) to ``fair'' (see NGC 1300,
 Figure~\ref{n1300}) to ``good'' (see NGC 4548,
 Figure~\ref{n4321.n4548}).  Table~\ref{table_frequency} lists the
 number and percentage of galaxies with clear detections of feathers
 for each of the three PDL classifications.  We see that the presence
 of delineated PDLs and feathers are highly correlated: 83\% of galaxies with
 good PDL delineation show feathers.  However, not
 all galaxies with delineated PDLs have feathers.  NGC 4450 is an
 example of a galaxy with delineated PDLs categorized as good which
 has no feathers.  Interestingly, NGC 4450 is also classified as an anemic
 galaxy (Elmegreen et al. 2002).

 The feathers detected in the 45 galaxies mentioned above
 originate at delineated PDLs.  We also find features resembling feathers present in
 flocculent galaxies that do not satisfy our strict definition of feathers 
 because they do not appear to originate at a PDL
 (as well as in four flocculent galaxies with feathers that
 clearly do originate at PDLs).  We detected such
 ``flocculent feathers'' in 17 out of 24
 flocculent galaxies.  We do not include cases of flocculent feathers without
 PDLs in Table 2.

 Also, within the central kiloparsec of some galaxies the images reveal fine,
 smaller scale extinction features.
 These interesting features, seen for example in NGC 5194 (Figure~\ref{m51.no.overlay}), 
 also do not appear to originate
 at a PDL and therefore we do not classify them as classic feathers.  Occasionally, the extinction
 features present
 could be interpreted as lattice structure.
 
 In general, from our sample of 223, those galaxies in which we did not definitively detect
 feathers fall into one or more of the following categories: a) the
 galaxy image is of lesser quality, typically due to exposure time or
 distance of the galaxy; b) PDLs are not clearly detected, which in
 some cases may be due to the quality of the observations; c) the
 galaxy is a flocculent with no clear PDLs; or d) the spiral arm
 structure is too complex or confusing, sometimes as the result of
 interaction.

 \subsection{Feather Frequency and Characteristics as a Function of Galaxy Type}
 Previous observational work emphasized that feathers and spurs
 are a characteristic of late-type spirals (e.g. Lynds 1970; Elmegreen
 1980).  Figure 17 shows that the sample of 223 spirals is sufficiently
 large to investigate feather frequency for all types of spirals.  We see that
 feathers are most common in Sb--Sc galaxies, with feathers clearly
 detected in 26--33\%.  
 
 At earlier Hubble types, the detection frequency is lower: 4/24
 or 17\%\ for Sab and 1/28 for Sa.  Clearly, feathers are quite common
 in somewhat earlier types (e.g. Sb) than previously thought.  For
 example, NGC 4736 (Sab) shows particularly clear feathers, as seen in
 Figure~\ref{n4579.n4736}. Still, feathers are definitely rarer at the
 earliest types.  This may be in part because some early type galaxies
 are relatively deficient in gas, or because the shallower pitch angle
 of the arms results in lower shock compression of the PDL.  Both of
 these might result in weaker PDLs; indeed, well-defined PDLs are
 detected in none of the 28 Sa galaxies and only 4 of the 24 Sab
 galaxies, compared to 31 of 129 Sb--Sc galaxies.

 At later Hubble types, there are no feathers detected (0/37 for Scd
 and 0/15 for Sd).  Note that no well-defined PDLs are detected in
 the 47 Scd--Sd galaxies,
 presumably because the spiral density enhancement
 is relatively weak in these generally lower luminosity galaxies, and
 this likely accounts for the paucity of feathers.


 We found little difference between barred and unbarred galaxies in
 the frequency of feather detection.  In particular, the frequencies
 are 14/66 (21\%) for SA, 17/78 (22\%) for SAB, and 14/73 (19\%) for
 SB.  As will be discussed in \S\,\ref{catalog}, parallel PDLs and
 ``lattice'' feather structure connecting the parallel PDLs are common
 in the arms of barred spirals.  
 
 We expected feathers to be associated with grand design spirals.  We
 were surprised to find feathers and PDLs in some flocculent spirals
 as well, since flocculents appear ``fleecy'' at optical wavelengths and
 typically lack large-scale, continuous spiral structure.  This
 absence of large-scale, continuous spiral structure is generally
 credited to an absence of spiral density waves within the galaxy.
 However, Thornley (1996) presented K' (2.1\micron) observations of
 four nearby flocculent spiral galaxies\footnote{NGC 2403, NGC 3521,
   NGC 4414, NGC 5055} that clearly show low-level spiral structure,
 which suggests that kiloparsec-scale spiral structure is more
 prevalent in flocculent galaxies than previously thought.
 Interestingly, a comparison of the two PDLs with feathers seen in
 the HST WFPC2 image of NGC 5055 to the stellar spiral arm observed at
 2.1\micron\ shows that the feather PDLs and stellar arms are related
 structures (see Figure~\ref{Thornley}).  This correlation between
 PDLs and spiral density waves suggests that feathers originating at
 delineated PDLs in flocculent galaxies may indicate an underlying
 spiral density wave not dominant at visible wavelengths.

 As described in \S\,\ref{frequency}, flocculents also harbor extinction structures which appear
 morphologically nearly identical to classic feathers, which we called flocculent feathers; these differ in
 that they are {\it not} associated with PDLs and tend to be associated
 with less star formation per feature.   
 They are nicely illustrated in NGC 7217
 (Figure~\ref{n6890.n7217}) as well as the parts of NGC 5055 (Figure
 12) not associated with the PDLs.  The number of flocculent feathers
 in a flocculent may greatly outnumber the classic feathers associated
 with PDLs.  Flocculent feathers are likely due to gaseous/dusty condensations sheared by differential
 rotation.  As mentioned
 previously, we detected
 flocculent feathers in 17 of 24 flocculent galaxies.


 \subsection{A Catalog of Feather Morphology}\label{catalog}
 
 General properties of the 45 galaxies with clearly delineated
 feathers are listed in Table~\ref{table_galaxies}.  See
 Figures~\ref{n0214.n1241.n1365}--\ref{n6890.n7217} for images of 21
 of the 45 galaxies with classic feathers.  We subjected this
 ``catalog'' of 45 galaxies to more detailed examination of feather
 morphology and characteristics.

 \subsubsection{``Beads on a String''}
 As noted above, feathers
 are typically associated with star formation, particularly near where they emerge from
 the PDL and outer edge of a spiral arm.  In 38 of the 45 galaxies with feathers, a
 series of bright OB associations and {\HII} regions occur which
 fit the ``beads on a
 string'' description used by Piddington (1970) and Elmegreen (1980).  The 38 
 galaxies with this characteristic are noted in Table~\ref{characteristics}.
 Typically, the star formation that comprises the ``beads'' is associated with feathers: near the 
 beginning of feathers, along the PDL; lining feathers, within the arm; and in the interarm, as 
 components of spurs.
 Spurs themselves are observed as short chains of star formation jutting outward from spiral arms
 at large angles.  For example, along the outer edge of the spiral
 arms in NGC 3631 (Figure~\ref{n3631.n4254}) several regions of OB
 associations are aligned in chains that are the beginnings of interarm spurs
 Also present in the southern arm of NGC 3631 is
 a chain of OB associations along the PDL.


 \subsubsection{Evolution of Feathers to Spurs}
 In eleven galaxies, we observe spurs on the
 outside of spiral arms, in the interarm regions, in addition to the star formation seen along 
 feathers within arms.  Within these eleven galaxies (see Table~\ref{characteristics})
 we find feathers that transition to spurs across the outer edge of the spiral
 arm (Figure~\ref{m51.no.overlay}), forming a composite feature.  These feathers appear to
 evolve into spurs with star formation developing as gas flows downstream from the PDL.  
 B88 and KO both predicted that spurs result from
 feathers.  KO's simulations follow the evolution of feathers through
 their fragmentation into self-gravitating clumps on the outer edge of
 a spiral arm.  The observations are consistent with
 this picture.

 \subsubsection{Elongated Feathers}
 As mentioned in \S\,\ref{char}, we also observe many feathers which are
 not solely contained within the luminous part of the spiral arms.
 These extend into the interarm similar to spurs; presumably these
 elongated feathers are the gaseous counterpart to spurs.  Twenty-nine
 galaxies listed in Table~\ref{characteristics} have elongated
 feathers, which are sometimes dotted
 with {\HII} regions.  Typically, elongated feathers are swept back
 more sharply in the interarm regions than they are at the
 outer edge of the luminous arm.  The decrease in pitch angle along
 elongated feathers is due to the large interarm shear referred to 
 in \S\,\ref{shape}.
 
 In many cases,
 elongated feathers merge
 into a common interarm dust lane.  In 15 galaxies, the farthest
 reaching feathers actually merge with the PDL of the next arm.


 \subsubsection{Lattice Structure}\label{lattice}
 Another intriguing feather structure, present in 14 galaxies noted in
 Table~\ref{characteristics}, is a series of feathers that appear
 to link one or more pairs of PDLs, forming a lattice within a single spiral arm.  An example of
 such a network of extinction features is in the northern arm of NGC
 4579 (Figure~\ref{n4579.n4736}) which includes several tiers
 of dust lanes connected by feathers emerging from the previous row.
 These multiple tracks should not be confused with the cases of elongated
 feathers that span an entire
 interarm region and merge with the PDL of another arm.  Lattices are most common in strongly barred 
 galaxies, and absent in unbarred galaxies.  In total there are nine
 SB galaxies and five SAB galaxies with lattices in one or more arms.  B88 predicted that multiple 
 tracks of feathers could
 form if the background flow was sufficiently unstable, allowing
 growth along both preferred directions, parallel and perpendicular to
 the spiral arm.

 \subsubsection{Shape of Feathers}\label{shape}
 The shape of a feather is a measure of the change in shear as
 gas passes through an arm.  The pitch angles of feathers, as mentioned in \S\,\ref{char}, vary
 with the phase of an arm.  Near the outer edges of an arm, feathers are almost always predominately
 trailing features.  Well-delineated
 cases show, however, that they originate as leading features from the PDL, 
 which marks the inner edge of a spiral arm.  


 
 The feathers described by B88, as well as those evident in the KO
 local MHD and DB hydrodynamical simulations, have the same
 characteristic shape as observed feathers.  In these cases, both observations and models,
 the feathers emerge at a large, leading angle with
 respect to a spiral arm, and the pitch angles
 decrease outward.  The outer portions of
 feathers are curved back into trailing features presumably by strong interarm
 shearing.
 
 Nearest to the spiral arm, the feathers produced in the WK hydrodynamic simulations curve in the
 opposite sense to observed feathers (from trailing to leading).  Further downstream in the interarm
 region, however, their features become trailing.
 This curvature is likely due to the rising background rotation curve they adopt.\footnote{If $\Omega$ is the local angular velocity, local wavevectors rotate counterclockwise wherever \[\frac{\partial \ln \Omega}{\partial \ln R}>0\] and clockwise where it is less than zero.  Since \[\frac{\partial \ln \Omega}{\partial \ln R} = \left( 1 + \frac{\partial \ln V_{c}}{\partial \ln R} \right)\frac{\Sigma}{\Sigma_o}~-2\] where $\Sigma$ is the local density and $\Sigma_o$ is the mean density, a solid body rotation curve would imply trailing-to-leading curvature in any overdense ($\Sigma > \Sigma_o$) region.}
 Further, although the arm-interarm
 density ratio is \(\sim\) 100, their models do not include
 self-gravity.  The WK model assumes physical
 conditions rather different from those where feathers have been observed: the gaseous disks are
 differentially rotating, self-gravity is important, and the
 arm-interarm surface density ratios are $\approx$ 10.  Thus the existing
 WK models are probably not relevant for observed feathers.


 \subsection{Feathers in the Spitzer 8\micron\ Band}
 
 We have discussed feathers as extinction features.  Despite the
 identification of some elongated feathers and also the association
 with spurs discussed in \S\,\ref{catalog}, extinction features can be
 difficult to trace where the background stellar density is low or
 where there is confusing foreground or embedded emission.  Recently
 released Spitzer SINGS data (Kennicutt et al. 2005) enable us to
 examine nine galaxies\footnote{NGC 0628, 1566, 4254, 4321, 4579,
   4725, 4736, 5055, and 5194} from our catalog in the IRAC 3.6--8\micron\ 
 bands.  These galaxies appear dramatically different at 8\micron\ from their
 appearance at optical wavelengths. 
 We observe that their 8$\mu$m spiral arms 
 have prominent, bright features that extend well into
 the interarm, referred to as ``filaments'' in NGC 5194 by Calzetti et al. (2005). 
 Particularly striking are NGC 5194 and NGC 0628 
 (see Figure~\ref{n5194.multi} and Figure~\ref{n0628.multi} 
 for IRAC 8\micron\ images).  

 The bright interarm features seen at 8\micron\ emerge from the
 spiral arms at large pitch angles.  Downstream from the arms, their pitch
 angle decreases with distance from the
 spiral arm.  Some of
 the 8\micron\ features span the entire interarm region and a few eventually merge
 with an outer arm, similarly to observed elongated feathers.
 The dark lines overlaid
 on the 8\micron\ images in Figure~\ref{n5194.multi} and
 Figure~\ref{n0628.multi} represent the location and length of feathers
 as measured in the HST image of the corresponding galaxy.  There appears
 to be a correlation between the elongated feathers observed in the
 visible and the bright interarm regions seen at 8\micron.  

 The primary source of emission at 8\micron\ is thought to be polycyclic aromatic
 hydrocarbons (PAHs), which are excited nonthermally by single UV
 photons (Sellgren 1984).  PAHs are found in diffuse atomic clouds and
 in photodissociation-regions (PDRs) surrounding molecular clouds,
 where they are excited or photodissociated by stellar UV radiation
 (e.g.
 van Dishoeck 2004).  Since the extinction that makes feathers visible is due to dust associated 
 with relatively dense interstellar gas, it is not surprising that feathers should be traced
 by PAH emission at 8\micron\, as we observe.

 \subsection{Association of Feathers with Molecular Gas}

 As discussed in the previous section, 8$\mu$m emission can be
 used to trace feathers into the interarms, where extinction features
 are more difficult to observe.  However, it is difficult to extract
 reliable gas column density estimates from either dust extinction or 8$\mu$m emission.  
 Dust extinction depends on the unknown distribution
 of dust relative to illuminating stars.  How are the stars distributed
 above and below the dust clouds?  Is the dust layer homogeneous or
 clumped?  What is the dust emissivity and the dust to gas ratio?
 Quantitative extraction of gas column density from 8$\mu$m emission
 has some of the same limitations, and also requires understanding of
 heating, formation, and destruction of PAHs.  By contrast,
 interpretation of CO 1-0 emission is perhaps more straightforward,
 although it too has limitations.  In this section, we evaluate the
 association of feathers with molecular gas emission as traced by CO 1-0
 emission.  Good observations are available for several galaxies from
 the BIMA SONG catalogue (Regan et al. 2001; Helfer et al. 2003).  
 
 In Figures 15 and 16 we show the locations of feathers (traced by
 lines) overlaid on maps of velocity-integrated CO 1-0 emission in NGC
 0628 and NGC 5194.  The figures show remarkable agreement between the
 location of CO peaks and the intersection of feathers with their PDL.
 Most feathers are very clearly associated with CO peaks and nearly all other feathers are plausibly 
 associated with CO peaks.  In fact, along the arms the correspondence is nearly one to one:
 nearly all CO peaks are associated with feathers.  It should be noted that
 initially we drew the feather lines without reference to the CO maps
 and later noted the close association.  For a few feathers, the
 precise location of the feather becomes unclear near the PDL owing to
 confusion from emission from massive star-forming regions.  In these
 cases, we found that a straight-line extrapolation of the
 feather inward intersected the PDL slightly inward from
 the nearest CO peak; if instead we assumed a curvature
 similar to neighboring feathers, the line intersected the PDL close
 to the CO peak.  For this small subset of feathers, we drew the line
 to have similar curvature and to intersect the CO peak.
 
 The  coincidence between  CO peaks  and the  ``base'' of  feathers is
 interesting,  because it  associates  feathers with  the highest  gas
 surface  density concentrations  in  the galaxy  disk.  It  therefore
 connects the feather phenomenon with much of the star formation in the galaxy disk.
 
 For some of the CO peaks, we see weak extensions in CO emission toward the
 outside of the arm along the feather.    The dynamic range of the CO map is limited;
 nonetheless for the stronger CO peaks it is clear that the gas column
 density of feathers rapidly decreases by at least a factor of 5--10
 with distance from the PDL.  Thus, contrary to the
 impression given by both the dust extinction maps and the 8$\mu$m
 Spitzer maps, CO maps show that the gas column density of feather features is by far the
 highest at the PDL.  Although it is conceivable that the CO
 emissivity is higher in the arm, it is unlikely to vary by a
 factor of ten.  The large enhancement in feather column density near the arm
 clearly pinpoints the PDL as the point of origin of the feather.
 Shear and
 divergent flow stretch out the condensations as the gas flows out of
 the arm.  While the molecular gas is strongly peaked at the PDL, star formation (as
 traced by H$\alpha$\ and 24\(\micron\)) tends to be more distributed along
 the feather \textit{downstream} of the PDL.
 




 \subsection{Feather Spacing and Gas Surface Density}\label{spacing}

 It is clear from inspection of the HST images that the spacing between
 feathers tends to increase with galactocentric radius, as can be seen
 for example in Figures 15 and 16.  This might be related to the gas
 surface density, which generally decreases with distance from the
 nucleus, and so it is interesting to compare  feather 
 spacings with gas column density. We made this comparison for two
 galaxies with both extensive feathers and good CO maps, NGC 0628 and
 NGC 5194.  For this comparison we used the BIMA SONG CO 1-0
 velocity-integrated maps, and sampled the CO maps along the PDLs of
 each spiral arm at 3\arcsec\ spacings (half beam width).  We then
 estimated the H$_2$ column densities using the relation

  \[N_{\rm H_{2}} = 2.2 \times 10^{20}\left(\frac{I_{\rm CO}}{\rm K~ km~s^{-1}}\right)~{\rm cm}^{-2},\]  
  (Strong et al. 1988).  The corresponding molecular surface density at
  each point is then

 \[\Sigma_{\rm H_{2}} = 2.17 \times 10^{-20}N_{\rm H_{2}}\cos i~~M_{\sun}~{\rm pc}^{-2}.\]

\noindent
 Noting that the gas surface densities are significantly lower in NGC 0628 than NGC 5194, we 
 included azimuthally averaged {\HI} data (Shostak \& van der Kruit 1984) to obtain a lower limit
 on the total gas surface density.  The radial atomic
 surface density is

 \[\Sigma_{\rm H_I} = 1.08 \times 10^{-20}N_{\rm H_I}\cos i~~M_{\sun}~{\rm pc}^{-2}\]
and the \textit{total} gas surface density is \(\Sigma_{\rm gas} = \Sigma_{\rm H_{2}} + \Sigma_{\rm H_I}~~M_{\sun}~{\rm pc}^{-2}.\)

 We plot (see Figures~\ref{m51.surfden} and
 \ref{n0628.surfden}) the gas surface density versus the \textit{deprojected}
 distance along the arms in both galaxies using the HST ACS image of
 NGC 5194 (larger FOV than WFPC2) and the HST WFPC2
 image of NGC 0628.  Also shown at the bottom of each figure are
 ticks marking the location of feathers along the arms.
 
 In NGC 5194, the surface density rises from the
 beginning of both arms, peaking near a galactocentric radius of 1 kpc.
 Thereafter, the surface density generally declines, although there
 are significant fluctuations.  Note that in the inner 1--2 kpc the
 peaks in Arm 1 tend to be higher (sometimes by a factor of two), but
 at larger radii Arm 2 peaks sometimes have higher surface density.
 Overall, it can be seen that the spacing between feathers increases
 as the surface density decreases along each arm, and that the feathers
 generally coincide with the gas column density peaks.
 
 NGC 0628 also exhibits a general decline in surface density with
 radius although there are significant peaks at larger radii.  As in
 NGC 5194, generally feather spacing increases and gas column density
 decreases as one moves out along the arm.  However, the gas surface
 densities are lower by at least a factor of three compared to
 NGC 5194; the peak surface density in NGC 5194 is at least an order
 of magnitude higher.

 The increase in feather spacing with decreasing gas surface density
 along a spiral arm and the association of feathers
 and gas surface density peaks evident in Figures~\ref{m51.surfden} and
 \ref{n0628.surfden} suggest a gravitational instability, such as the
 Jeans instability.  To investigate this further, we estimated the
 Jeans length as

 \[\lambda_{Jeans} = c_{s}^{2} / (G\Sigma),\] 
 where \(c_{s}\) is the effective sound speed.  
 We adopt
 \(c_{s}\) = 7 km~s\(^{-1}\) for both NGC~0628 and NGC 5194.  
 
 Figure~\ref{m51.avg} shows the feather separations and Jeans lengths,
 along with their ratio as a
 function of distance along each arm in NGC 5194.
 Figure~\ref{n0628.avg} shows the same quantities for NGC 0628.  
 Noticing that feathers tend to ``clump'' together in groups, with
 larger distances separating the groups than the average distance
 between the feathers within a group, we averaged the separations, Jeans length, and ratios
 separately for each group. For NGC 5194, it can be seen that the feather spacing increases from
 $\sim$200 pc in the inner galaxy to 500--1000 pc in the outer galaxy.  Since the Jeans length 
 increases by a similar factor, the ratio of separation to Jeans length does not exhibit a radial
 trend.  NGC 0628 also shows evidence for some increase in feather spacing with galactocentric 
 radius, while the ratio of separation to Jeans length is relatively flat.


 While the feather spacings in NGC 0628 and
 NGC 5194 are roughly similar, the measured ratio of feather spacing to Jeans length
 is a factor of 4--6 higher in NGC 5194 and differs between its arms.  The lower ratios
 in NGC 0628 are probably due in part to the greater fraction of gas in NGC 0628 likely being 
 in atomic form.  Although we have included azimuthally averaged {\HI} 
 (Shostak \& van der Kruit 1984)
 to calculate the total gas surface density, the {\HI} data are at low resolution 
 (13.8\arcsec x 48.5\arcsec), and therefore the column densities are lower limits.  In NGC 5194, the
 addition of {\HI} is not a problem 
 since the 
 gas is overwhelmingly molecular,
 but in NGC 0628 the unresolved {\HI} column density could be substantial.  Another factor 
 lowering the measured
 ratio is that a significant fraction of the CO flux is likely undetected in NGC 0628, owing to low
 signal to noise (Helfer et al. 2002, 2003).  Thus, while the peaks  and radial trends observed for 
 NGC 0628 are credible, the magnitudes of the Jeans length and ratio are probably only useful
 for NGC~5194.  
 

 We also measured the separation of feathers in NGC 3433 and NGC 5985, 
 Figure~\ref{n2339.n3177.n3433} and Figure~\ref{n5055.n5985}.  As in NGC 0628 and NGC 5194, 
 both galaxies display a trend of 
 increasing separation along an arm.  In NGC 3433, the average
 separation
 along Arm 1 is 310 pc between 0.9 kpc and 2.9 kpc galactocentric radius.  For Arm 2, the average 
 is 380 pc between 1.5 kpc and 1.6 kpc and 430 pc between 2.7 kpc and 7.1 kpc.  In NGC 5985, the 
 average separation along Arm 1 is 445 pc between 3.7 kpc and 5.1 kpc.  For Arm 2, the average is
 770 pc between 3.0 kpc and 3.1 kpc.  

 It would of course be interesting to measure and characterize the typical feather spacing in more
 galaxies.  However, either due to the quality of observations, the complexity of extinction 
 substructure, or the contamination due to star formation, most of the images did not lend 
 themselves to reliable estimates of feather spacings.
 As we have shown, molecular gas offers a 
 non-extincted, less-confused tracer of
 feathers.  Therefore, separation measurements using CO observations with CARMA and ALMA may yield 
 better estimates in more galaxies.


 Feather spacings found in NGC 0628 and NGC 5194 are typically larger than the 100--200 pc range
 predicted by WK.  DB predicted feather spacings of 700 pc; such large spacings are only seen in 
 the outer part of Arm 2 in NGC 5194.  We note that DB and WK did not include self-gravity in 
 their calculation;  particularly in NGC 5194, the observed gas densities are comparable or larger 
 the stellar surface densities, indicating that self-gravity is likely important.

 KO and Kim \& Ostriker 2006 predicted the characteristic spacing of feathers to be 
 up to 7--10 times the local Jeans length for
 their magnetized models, which are stable
 to quasi-axisymmetric modes.  For models which are unstable to quasi-axisymmetric modes, the arms 
 do not have a well-defined ``pre-fragmentation'' surface density, but the feather spacings are
 in general smaller.  The range in ratios of feather spacing to Jeans length observed in 
 NGC 5194 centers on the values of 7--10 predicted by KO.  This, along with the increase in feather 
 spacing with decreasing gas surface density and the correlation between the location of feathers and gas surface density peaks, is  consistent with the view that feathers form via a 
 gravitational instability.

 \section{Conclusions}

 We find extinction feathers
 in nearly 20\% of 233 spiral galaxies.  We show that feathers are most common in
 Sb--Sc galaxies; Sb--Sc galaxies in which we did not detect feathers either had
 poor quality images, or flocculent or complex structure.  Feathers are rare in Sa galaxies and undetected
 in Scd--Sd galaxies.  The presence of feathers is closely tied to the existence of a primary dust lane
 (PDL).  The probability of detecting feathers increases with PDL delineation; the highest being 83\% for 
 spirals with
 well-delineated PDLs, within which feathers are ubiquitous.  Characteristically, feathers: (1)
 are associated with bright star forming regions within spiral arms and
 interarm regions; (2) extend beyond the outer edge of spiral arms, sometimes far
 into interarm regions, and merging with the PDL of
 another arm; (3) transition or evolve into
 stellar spurs; and (4) often form lattice structures or multiple rows of
 feathers within a single spiral arm in barred galaxies.  Furthermore, we find that 
 the spacing of feathers is related
 to the molecular surface density along spiral arms; (1) typically, the
 distance between feathers increases as the molecular surface density
 decreases and (2) the majority of feathers originate in regions of
 higher gas surface density.  The
 mean separation of feathers is
 \(1.7 \lambda_{Jeans}\) and \(10.4 \lambda_{Jeans}\) in NGC 0628 and NGC 5194; the value for
 NGC 0628 is likely an underestimate due to poor signal to noise and a lower limit of {\HI} data.
 The above observable characteristics are consistent with models in which feathers are
 produced by local gravitational instabilities (i.e. Jeans or
 magneto-Jeans instability) in the gas.

 \section{Acknowledgments}
 This research is supported in part by grants AST-0228974 and
 AST-0507315 from the National Science Foundation.  We made use of the
 NASA/IPAC Extragalactic Database (NED) which is operated by the Jet
 Propulsion Laboratory, California Institute of Technology, under
 contract with the National Aeronautics and Space Administration.
 This work is based in part on observations made with the Spitzer
 Space Telescope, which is operated by the Jet Propulsion Laboratory,
 California Institute of Technology under NASA contract 1407.


 \clearpage

 \clearpage


 \begin{deluxetable}{lccc}

 \tablecaption{Delineation of Primary Dust Lanes and Frequency of Feather Detection}

 \tablewidth{0pt}

 \tablehead{\colhead{Primary Dust} & \colhead{\# of Galaxies} & \colhead{\# of Galaxies} & \colhead{\% with Feathers} \\ 
 \colhead{Lane Delineation} & \colhead{} & \colhead{with Feathers} & \colhead{} } 

 \startdata
 Poor & 142 & 4 & 3\% \\
 Fair & 46 & 12 & 26\% \\
 Good & 35 & 29 & 83\% \\
 \hline
 \hline
 Total & 223 & 45 & 20\% \\

 \enddata
 \label{table_frequency}
 \end{deluxetable}

 \clearpage


 \begin{deluxetable}{cccccccc}

 \tabletypesize{\footnotesize}
 \tablecaption{Properties of the 45 Galaxies with Clearly Delineated Feathers}
 \tablewidth{0pt}

 \tablehead{\colhead{Galaxy} & \colhead{Figure} & \colhead{Type} & \colhead{Incl.} & \colhead{PA} & \colhead{B\(_T\)} & \colhead{CZ} & \colhead{Arm Class} \\ 
 \colhead{} & \colhead{Number} & \colhead{} & \colhead{(deg)} & \colhead{(deg)} & \colhead{(mag)} & \colhead{(km s\(^{-1}\))} & \colhead{(EE)} }

 \startdata
 IC 2056 & & RSABRbc* & 34 & 8 & 12.48 & 1099 & \\

 NGC 0214 &~\ref{n0214.n1241.n1365} & SABRbc & 42 & 35 & 12.86 & 4495 & 9 \\

 NGC 0289 & & SBTbc & 45 & 130 & 11.72 & 1690 & 12 \\

 NGC 0488 & & SARb & 42 & 15 & 11.15 & 2233 & 3 \\

 NGC 0613 & & SBTbc & 41 & 120 & 10.73 & 1510 & 9 \\

 NGC 0628 &~\ref{n0628.overlay},~\ref{n0628.no.overlay} & SASc & 24 & 25 & 10.01 & 632 & 9 \\

 NGC 0986 & & SBTab & 41 & 150 & 11.64 & 1994 & \\

 NGC 1241 &~\ref{n0214.n1241.n1365} & SBTb & 53 & 140 & 12.64 & 3939 & 4 \\

 NGC 1300 &~\ref{n1300} & SBTbc & 49 & 106 & 11.11 & 1592 & 12 \\

 NGC 1365 &~\ref{n0214.n1241.n1365}  & SBSb & 57 & 32 & 10.23 & 1675 & 12 \\

 NGC 1512 & & SBRa & 51 & 90 & 11.13 & 735 & 6 \\

 NGC 1566 &~\ref{n1566.n2336} & SABSbc & 37 & 60 & 10.13 & 1449 & 12 \\

 NGC 1667 & & SABRc & 39 & & 12.77 & 4587 & \\

 NGC 2207 & & SABTbcP & 50 & 141 & & 2728 & 5 \\

 NGC 2336 &~\ref{n1566.n2336} & SABRbc & 57 & 178 & 11.26 & 2205 & 9 \\

 NGC 2339 &~\ref{n2339.n3177.n3433} & SABTbc & 41 & 175 & 11.98 & 2361 & 5 \\

 NGC 2997 & & SABTc & 41 & 110 & 10.06 & 1090 & 9 \\

 NGC 3177 &~\ref{n2339.n3177.n3433} & SATb & 36 & 135 & 12.9 & 1220 & 6 \\

 NGC 3433 &~\ref{n2339.n3177.n3433} & SASc & 27 & 50 & 12.29 & 2591 & 9 \\

 NGC 3631 &~\ref{n3631.n4254} & SASc & 17 &  & 11.05 & 1143 & 9 \\

 NGC 3783 & & PSBRab & 27 & & & 2926 & 9 \\

 NGC 4030 & & SASbc & 44 & 27 & & 1449 & 9 \\

 NGC 4254 &~\ref{n3631.n4254} & SASc & 29 &  & 10.17 & 2453 & 9 \\

 NGC 4303 & & SABTbc & 27 & & 10.18 & 1607 & 9 \\

 NGC 4321 &~\ref{n4321.n4548} & SABSbc & 32 & 30 & 10.26 & 1579 & 12 \\

 NGC 4394 & & RSBRb & 27 & & 11.73 & 772 & 6 \\

 NGC 4548 &~\ref{n4321.n4548} & SBTb & 37 & 150 & 11.04 & 498 & 5 \\

 NGC 4579 &~\ref{n4579.n4736} & SABTb & 37 & 95 & 10.68 & 1627 & 9 \\

 NGC 4593 & & RSBTb & 42 & & & 2662 & 5 \\

 NGC 4647 & & SABTc & 37 & 125 & 11.94 & 1421 & 3 \\

 NGC 4725 & & SABRabP & 45 & 35 & 10.11 & 1180 & 6 \\

 NGC 4736 &~\ref{n4579.n4736} & RSARab & 36 & 105 & 8.99 & 297 & 3 \\

 NGC 5055 &~\ref{n5055.n5985} & SATbc & 55 & 105 & 9.57 & 516 & 3 \\

 NGC 5194 &~\ref{m51.overlay},~\ref{m51.no.overlay}& SASbcP & 24 & 170 & 9.08 & 463 & 12 \\

 NGC 5236 &~\ref{n5236} & SABSc & 27 &  & 8.31 & 503 & 9 \\

 NGC 5248 & & SABTbc & 44 & 110 & 10.97 & 1189 & 12 \\

 NGC 5383 & & PSBTb*P & 32 & 85 & 12.05 & 2226 & 12 \\

 NGC 5427 & & SAScP & 32 & & 11.93 & 2645 & 9 \\

 NGC 5643 & & SABTc & 29 & & 10.74 & 1163 & \\

 NGC 5970 & & SBRc & 48 & 88 & 12.24 & 2063 & 9 \\

 NGC 5985 &~\ref{n5055.n5985} & SABRb & 58 & 13 & 11.67 & 2467 & 9 \\

 NGC 6753 & & RSARb & 29 & 30 & 11.97 & 3142 & 8 \\

 NGC 6814 & & SABTbc & 21 & & 12.06 & 1509 & 9 \\

 NGC 6890 &~\ref{n6890.n7217} & SATb & 37 & 152 & 13.05 & 2471 &  \\

 NGC 7392 & & SASbc & 54 & 123 & 12.62 & 2908 & 5 \\

 \enddata
\label{table_galaxies}

 \end{deluxetable}
 \clearpage


 \begin{deluxetable}{ccccccccccc}
 \tabletypesize{\footnotesize}

 \tablecaption{Feather Characteristics}
 \tablewidth{0pt}


 \tablehead{\colhead{Galaxy} & \colhead{Figure} &  \colhead{Feather\tablenotemark{a}} & \colhead{``Beads on} & \colhead{Elongated} & \colhead{Lattice} \\

 \colhead{} & \colhead{Number} &  \colhead{\(\Rightarrow\) Spur} & \colhead{a String''} & \colhead{Feathers} & \colhead{of Feathers}}

 \startdata
 IC 2056 & &  & x & & x \\

 NGC 0214 &~\ref{n0214.n1241.n1365} &   & x & x &  \\

 NGC 0289 & &   & & x & x\\

 NGC 0488 & &   & & x & \\

 NGC 0613 & &   & x & & \\

 NGC 0628 &~\ref{n0628.overlay},~\ref{n0628.no.overlay}  &   & x & x &  \\

 NGC 0986 & &  x & x & x & x  \\

 NGC 1241 &~\ref{n0214.n1241.n1365} &    & x &  &  \\

 NGC 1300 &~\ref{n1300} &   x & x & x & x \\

 NGC 1365 &~\ref{n0214.n1241.n1365} &    & x &  & x \\

 NGC 1512 & &   & x & x & x \\

 NGC 1566 &~\ref{n1566.n2336} &  x & x & x &  \\

 NGC 1667 & &   & x & x & \\

 NGC 2207 & &  & x & & \\

 NGC 2336 &~\ref{n1566.n2336} &    & x & x &  \\

 NGC 2339 &~\ref{n2339.n3177.n3433} &    &  &  &  \\

 NGC 2997 &  &  & x & & \\

 NGC 3177 &~\ref{n2339.n3177.n3433} &    & x & x &  \\

 NGC 3433 &~\ref{n2339.n3177.n3433} &   x & x & x &  \\

 NGC 3631 &~\ref{n3631.n4254} &  x & x & x &  \\

 NGC 3783 & &  & x & x & \\

 NGC 4030 & &    & x & & \\

 NGC 4254 &~\ref{n3631.n4254} &    & x & x &  \\

 NGC 4303 & &   & x & & \\

 NGC 4321 &~\ref{n4321.n4548} &    & x &  &  \\

 NGC 4394 & &   & & x & x \\

 NGC 4548 &~\ref{n4321.n4548} &   & x & x & x \\

 NGC 4579 &~\ref{n4579.n4736} &    &  & x & x \\

 NGC 4593 & &   & x & & x \\

 NGC 4647 & &   & x & x &  \\

 NGC 4725 & &   x & x & x & \\

 NGC 4736 &~\ref{n4579.n4736} &  & x & x &  \\

 NGC 5055 &~\ref{n5055.n5985}&   & & & \\

 NGC 5194 &~\ref{m51.overlay},~\ref{m51.no.overlay} &  x & x & x &  \\

 NGC 5236 &~\ref{n5236} &   x & x & x & x \\

 NGC 5248 & &   & x &  & \\

 NGC 5383 & &  & x & & x \\

 NGC 5427 & &   x & x & x & \\

 NGC 5643 & &   & x &  & x \\ 

 NGC 5970 & &   & x & x & \\

 NGC 5985 &~\ref{n5055.n5985} &   x & x & x & x \\

 NGC 6753 & &   & x & x & \\

 NGC 6814 & &   & & x & \\

 NGC 6890 &~\ref{n6890.n7217} &    & x &  &  \\

 NGC 7392 & &  & x & x & \\

 \enddata

 \tablenotetext{a}{Galaxies within which we identified feathers transitioning 
 into spurs.}

 \label{characteristics}

 \end{deluxetable}

 \clearpage



 \begin{figure}
 \centering
 \caption{A prototypical galaxy with feathers, NGC 5194.  Feathers are marked by white lines.}
 \label{m51.overlay}
 \end{figure}
 \clearpage

 \begin{figure}
 \centering
 \caption{A prototypical galaxy with feathers, NGC 0628.  Feathers are marked by white lines.}
 \label{n0628.overlay}
 \end{figure}

\clearpage

 \begin{figure}
 \centering
 \caption{NGC 5194, with the feather overlay omitted.}
 \label{m51.no.overlay}
 \end{figure}

 \clearpage

 \clearpage

 \begin{figure}
 \centering
 \caption{NGC 0628, with the feather overlay omitted.}
 \label{n0628.no.overlay}
 \end{figure}

 \clearpage


 \begin{figure}
 \centering
 \caption{{\it Top left}: NGC 0214, {\it Top right}: NGC 1241, {\it Bottom}: NGC 1365.  The white arrows mark identified feathers.}
 \label{n0214.n1241.n1365}
 \end{figure}

 \clearpage

 \begin{figure}
 \centering
 \caption{Image: NGC 1300.  The white arrows mark identified feathers.}
 \label{n1300}
 \end{figure}

 \clearpage

 \begin{figure}
 \centering
 \caption{{\it Top}: NGC 1566, {\it Bottom}: NGC 2336.  The white arrows mark identified feathers.}
 \label{n1566.n2336}
 \end{figure}

 \clearpage

 \begin{figure}
 \centering
 \caption{{\it Top left}: NGC 2339, {\it Top right}: NGC 3177, {\it Bottom}: NGC 3433.  The white arrows mark identified feathers.}
 \label{n2339.n3177.n3433}
 \end{figure}

 \clearpage

 \begin{figure}
 \centering
 \caption{{\it Top}: NGC 4254, {\it Bottom}: NGC 3631.  The white arrows mark identified feathers.}
 \label{n3631.n4254}
 \end{figure}

 \clearpage

 \begin{figure}
 \centering
 \caption{{\it Top}: NGC 4321, {\it Bottom}: NGC 4548.  The white arrows mark identified feathers.}
 \label{n4321.n4548}
 \end{figure}

 \clearpage

 \begin{figure}
 \centering
 \caption{{\it Top}: NGC 4579, {\it Bottom}: NGC 4736.  The white arrows mark identified feathers.}
 \label{n4579.n4736}
 \end{figure}

 \clearpage

 \begin{figure}
 \centering
 \caption{{\it Top}: NGC 5055, {\it Bottom}: NGC 5985.  The white arrows mark identified feathers.}
 \label{n5055.n5985}
 \end{figure}

 \clearpage

 \begin{figure}
 \centering
 \caption{Image: NGC 5236 (Larsen and Richtler 1999).  The white arrows mark identified feathers.}
 \label{n5236}
 \end{figure}

 \clearpage

 \begin{figure}
 \centering
 \caption{{\it Top}: NGC 6890, the white arrows mark identified feathers. {\it Bottom}: NGC 7217, an example of a galaxy with flocculent feathers.  }
 \label{n6890.n7217}
 \end{figure}

 \clearpage


 \clearpage

 \begin{figure}
 \centering
 \caption{\footnotesize{Two views of NGC 5194, with lines marking the feathers 
 from HST image shown in Fig.~\ref{m51.overlay} -- {\it Top}: Spitzer 8\micron\ image.  
 {\it Bottom}: BIMA SONG CO(J=1-0) image.}}
 \label{n5194.multi}
 \end{figure}

 \clearpage

 \begin{figure}
 \centering
 \caption{\footnotesize{Two views of NGC 0628, with lines marking the
 feathers from HST image shown in Fig.~\ref{n0628.overlay} -- {\it Top}: Spitzer 8\micron\ image.  
{\it Bottom}: BIMA SONG CO(J=1-0).}}
 \label{n0628.multi}
 \end{figure}

 \clearpage

 \begin{figure}
 \centering
 \plotone{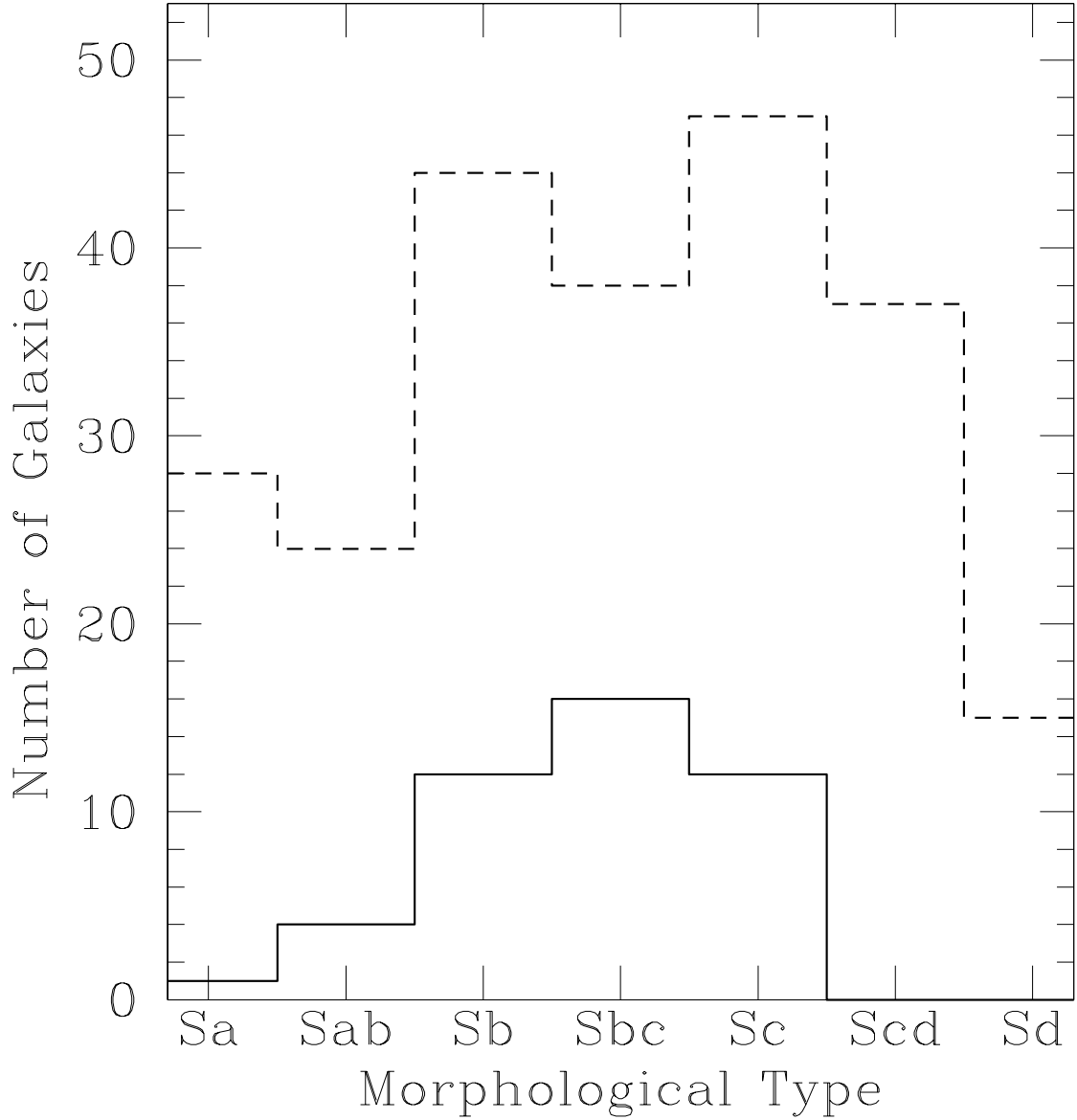}
 \caption{Distribution of galactic morphological types included in sample.  The dashed line represents the distribution of morphological types of all the galaxies in the sample.  The solid line shows the distribution for galaxies with feathers. }
 \label{morph.hist}
 \end{figure}

 \clearpage

 \begin{figure}
 \centering
 \caption{\small{{\it Top}: WFPC2 HST F814W band image of NGC 5055.  {\it Bottom}: 2.1\micron\ image.  The thin grey lines in the HST image indicate observed PDLs within the galaxy.  The same lines are overlaid on the 2.1\micron\ image, which show there is a clear association between PDLs observed in the visible and observed infrared emission marking the peak of the stellar arms.} }
 \label{Thornley}
 \end{figure}

 \clearpage

\clearpage

\begin{figure}
\plotone{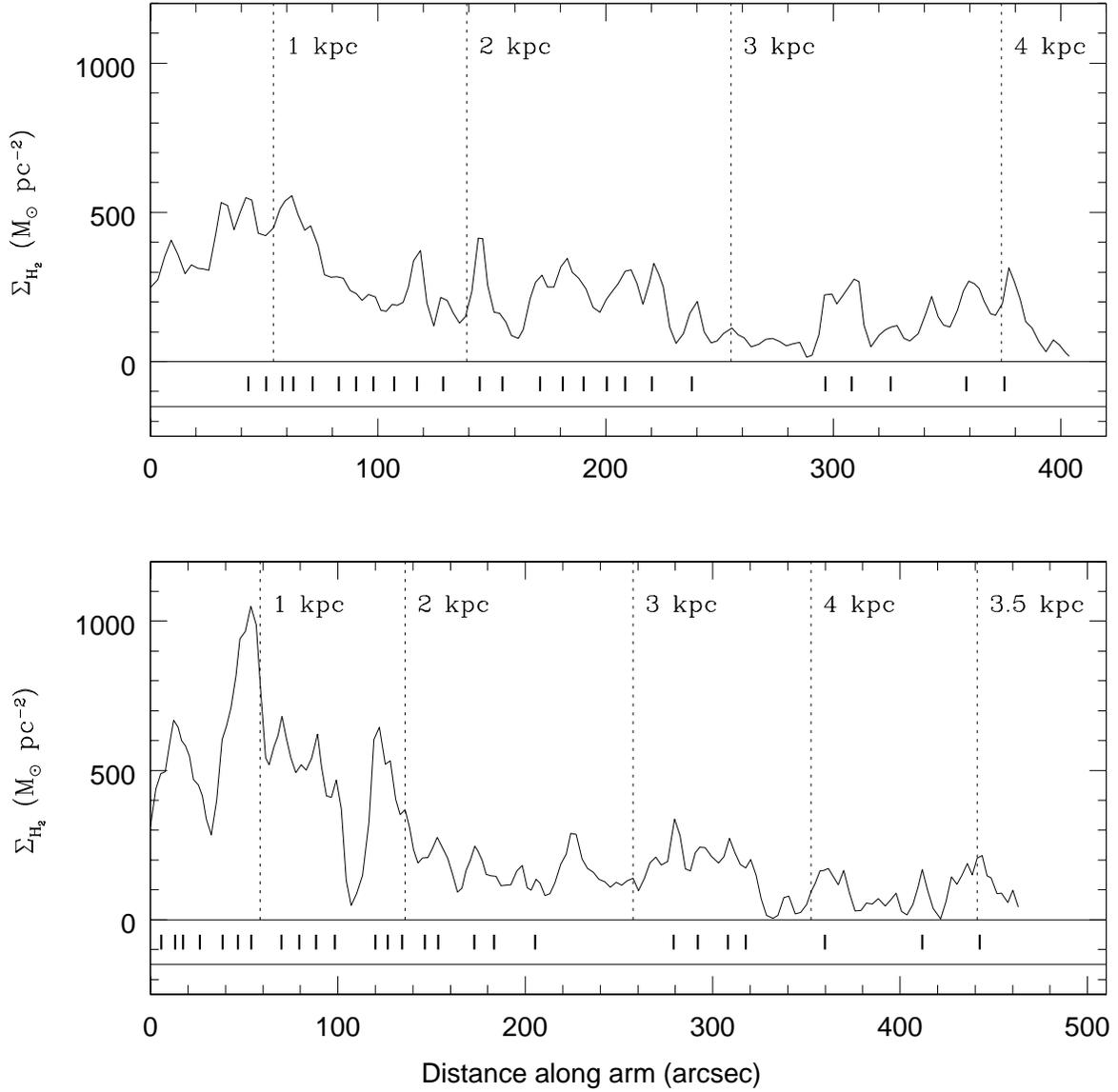}
\caption{Variation in surface density along Arm 1 (top) and Arm 2 (bottom) in NGC 5194.  The short vertical lines indicate the location of a feather along an arm.  The dotted vertical lines indicate galactocentric radius.  Note that Arm 2 bends inward near \(\sim\)4 kpc.}
\label{m51.surfden}
\end{figure}

\clearpage

\begin{figure}

\plotone{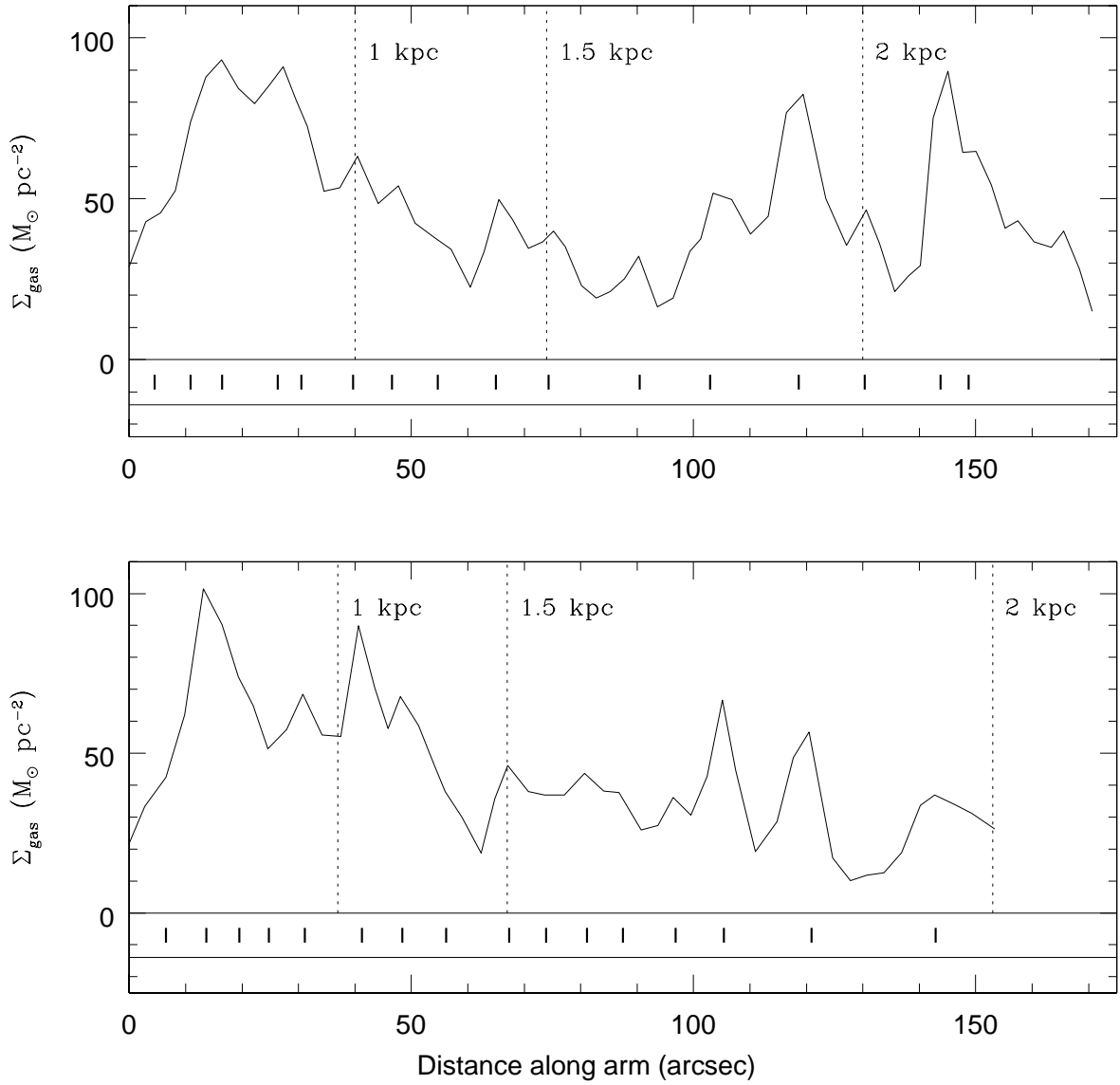}
\caption{The same as Figure~\ref{m51.surfden}, for NGC 0628.}
\label{n0628.surfden}
\end{figure}

\clearpage

\begin{figure}
\plotone{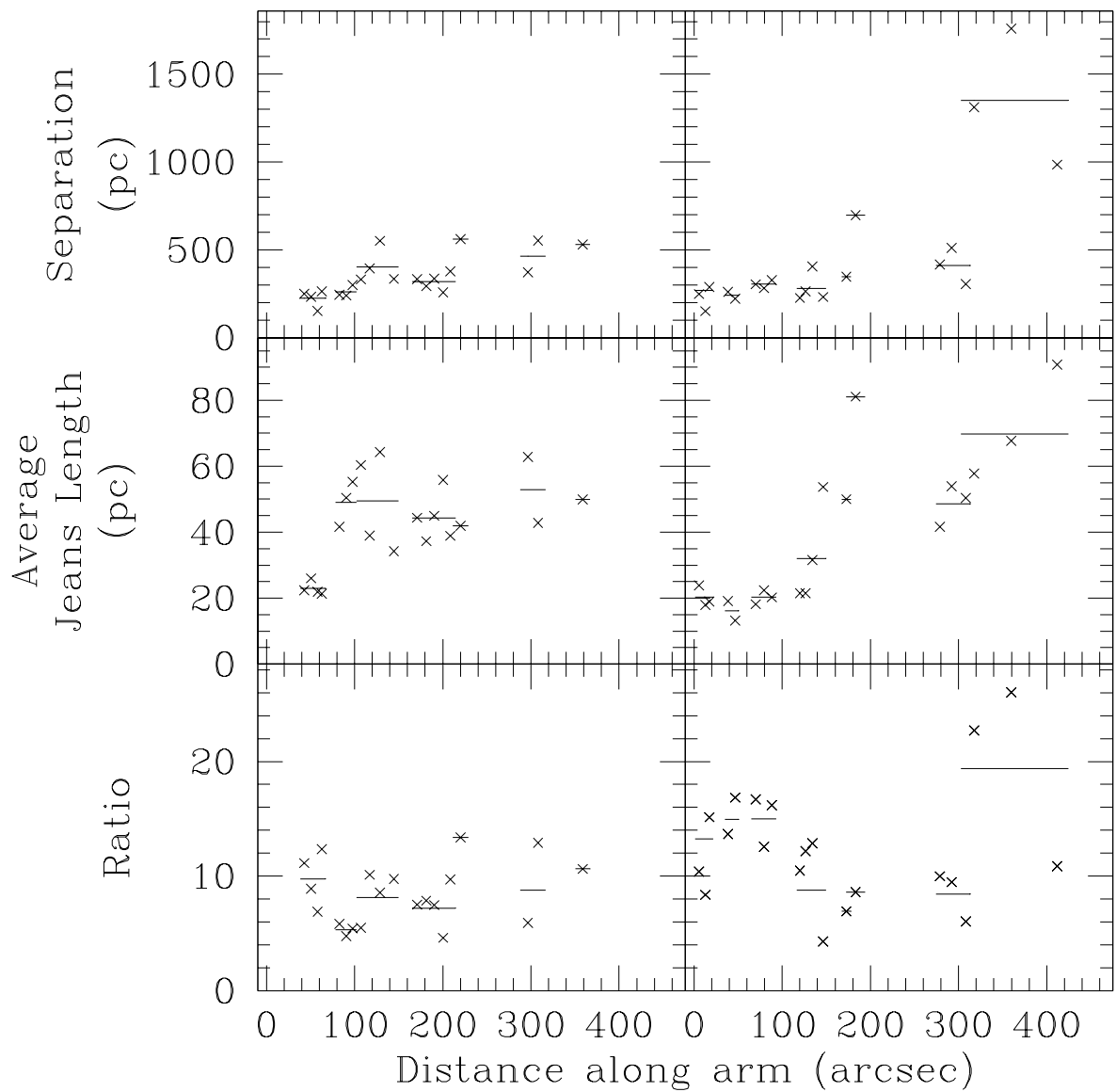}

\caption{
  Comparison of feather separation to the local Jeans length averaged
  over 6\arcsec\ in NGC 5194, as a function of distance along an arm;
  the left and right columns show data for Arms 1 and 2,
  respectively.  ({\it Top}) Deprojected feather separation.  ({\it Middle})
  Average Jeans length at each feather.  ({\it Bottom}) Ratio of deprojected
  feather spacing to the local Jeans length.  The lines indicate the mean value for each group of
  feathers along an arm; the length of a line indcates the width of a group.}
\label{m51.avg}
\end{figure}

\clearpage




\begin{figure}

\plotone{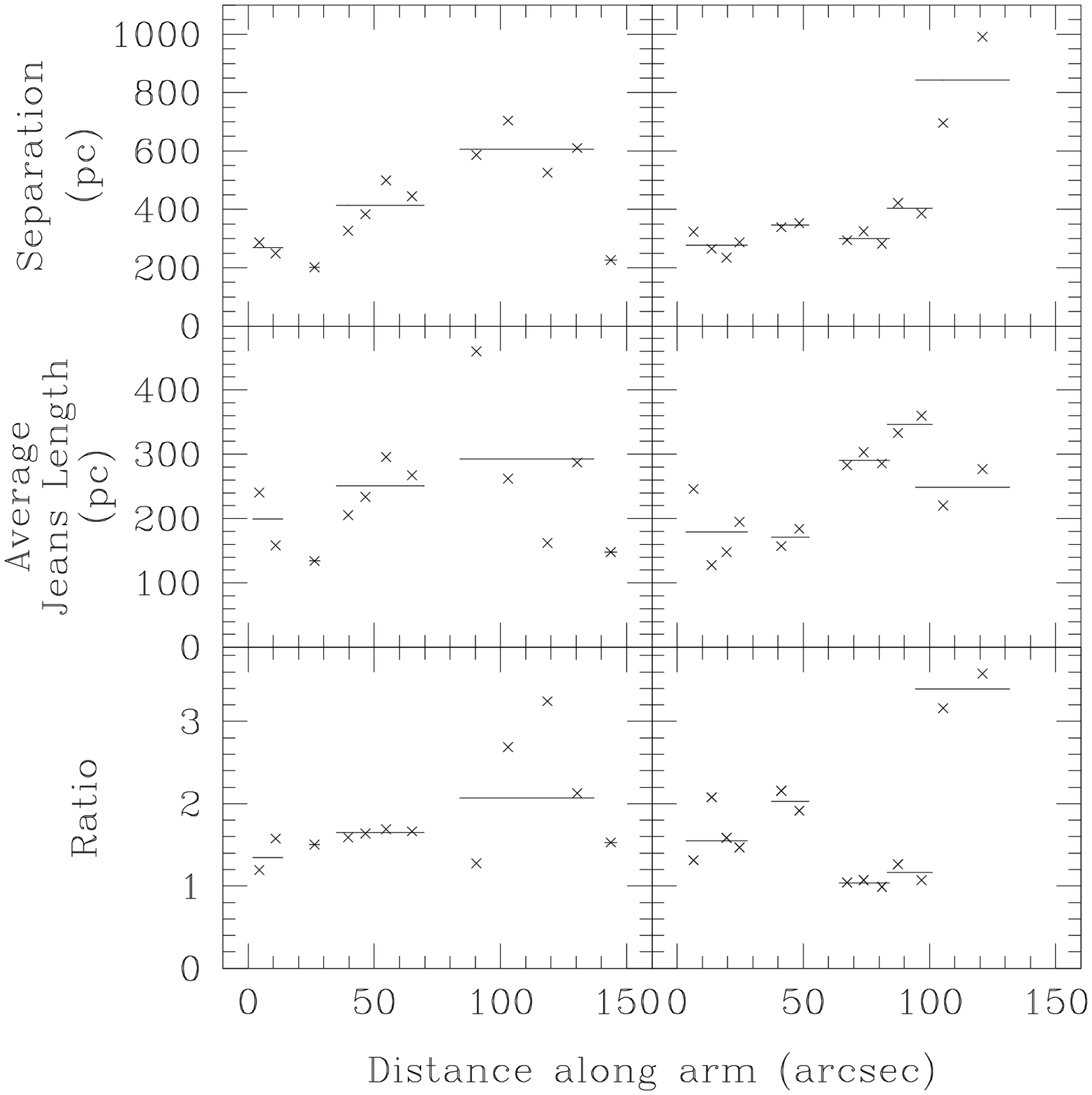}
\caption{The same as Figure~\ref{m51.avg}, for NGC 0628.}
\label{n0628.avg}
\end{figure}

\clearpage

\end{document}